# Crossover between weak localization and weak antilocalization in magnetically doped topological insulator

Minhao Liu[1,*], Jinsong Zhang[1,*], Cui-Zu Chang[1,2,*], Zuocheng Zhang[1], Xiao Feng[2], Kang Li[2], Ke He[2,†], Li-li Wang[2], Xi Chen[1], Xi Dai[2], Zhong Fang[2], Qi-Kun Xue[1,2], Xucun Ma[2], and Yayu Wang[1,†]

[1]*Laboratory of Low Dimensional Quantum Physics, Department of Physics, Tsinghua University, Beijing 100084, P. R. China*

[2]*Institute of Physics, Chinese Academy of Sciences, Beijing 100190, P. R. China*

\* *These authors contributed equally to this work.*

† Email: kehe@aphy.iphy.ac.cn; yayuwang@tsinghua.edu.cn

**Topological insulators (TI) are a new class of quantum materials with insulating bulk enclosed by topologically protected metallic boundaries[1-3]. The surface states of three-dimensional TIs have spin helical Dirac structure[4-6], and are robust against time reversal invariant perturbations. This extraordinary property is notably exemplified by the absence of backscattering by nonmagnetic impurities[7-9] and the weak antilocalization (WAL) of Dirac fermions[10-12]. Breaking the time reversal symmetry (TRS) by magnetic element doping is predicted to create a variety of exotic topological magnetoelectric effects[13-18]. Here we report transport studies on magnetically doped TI Cr-$Bi_2Se_3$. With increasing Cr concentration, the low temperature electrical conduction exhibits a characteristic crossover from WAL to weak localization (WL). In the heavily doped regime where WL dominates at the ground state, WAL reenters as temperature rises, but can be driven back to WL by strong magnetic field. These complex phenomena can be explained by a unified picture involving the evolution of Berry phase with the energy gap opened by magnetic impurities. This work demonstrates an effective way to manipulate the topological transport properties of the TI surface states by TRS-breaking perturbations.**

$Bi_2Se_3$ is an ideal three-dimensional TI due to its large bulk energy gap (~ 300meV) and a Dirac point located well outside the bulk bands[19,20]. On the surface of magnetically doped $Bi_2Se_3$ single crystals, Angle-resolved photoemission spectroscopy (ARPES) has revealed the opening of an energy gap at the Dirac point[21] and the creation of odd multiples of Dirac fermions[22]. The Chromium (Cr) doped $Bi_2Se_3$ films studied here are grown on sapphire substrate by molecular beam epitaxy (MBE), as shown schematically in Fig. 1a. The thickness of the films is kept at three quintuple layers (3QL) because in this ultrathin regime the conduction is dominated by the surface states[23].

Figure 1b shows the sheet resistance ($R_□$) vs temperature ($T$) for a pure $Bi_2Se_3$ and four Cr doped $Bi_{2-x}Cr_xSe_3$ films with $x$ = 0.04, 0.07, 0.13, and 0.23. The $R_□$–$T$ curves always show metallic behavior (positive slope) at high $T$ and insulating behavior (divergent increase) at low $T$. Fig. 1c shows that both the minimum resistance value and the onset temperature for the insulating behavior increase with $x$. This is the first indication that magnetic impurities play an important role in charge transport in TIs. The effect is more dramatic at low $T$, where the resistance of the $x$ = 0.23 film is enhanced by more than two orders of magnitude from that of pure $Bi_2Se_3$. Evidently the electron conduction capability of the Dirac fermions is severely damaged by magnetic impurities.

Figure 2a to 2e display the variation of Hall resistance ($R_{yx}$) with magnetic field ($H$) for the five $Bi_{2-x}Cr_xSe_3$ films. The negative $R_{yx}$ in all samples indicates that the Fermi energy ($E_F$) always lies above the Dirac point. For pure $Bi_2Se_3$ (Fig. 2a), the field dependence of $R_{yx}$ measured at the base temperature $T$ = 1.5K is a straight line, as expected for nonmagnetic conductors. For magnetically doped films, the low $T$ Hall traces start to develop a curvature at weak field. The nonlinearity becomes more pronounced as $x$ increases, apparently due to the larger concentration of doped magnetic moments. The overall pattern of the Hall curves in the heavily doped $x$ = 0.23 sample (Fig. 2e) resembles the anomalous Hall effect (AHE) in ferromagnetic materials, in which the total Hall resistance can be expressed as

$$R_{yx} = R_A \cdot M(T, H) + R_N \cdot H.$$

Here $R_A$ is the anomalous Hall coefficient, $M$ is the magnetization of the sample, and $R_N$ is the normal Hall coefficient. The strength of the anomalous term, hence the magnitude of the magnetization, can be estimated by the intercept $R_{yx}^0$ obtained by extrapolating the high field linear Hall curve to zero field, as shown by the broken line in Fig. 2e. The solid squares in Fig. 3a summarize the $R_{yx}^0$ value of the five $Bi_{2-x}Cr_xSe_3$ films measured at $T = 1.5K$, which increases systematically with the magnetic impurity concentration. The solid squares in Fig. 3b display the $R_{yx}^0$ value of the heavily doped $x = 0.23$ sample as a function of temperature. The decrease of $R_{yx}^0$ with increasing $T$ is consistent with the reduction of magnetization by thermal fluctuations.

Figure 2f to 2j display the magnetoconductance (MC) $\delta\sigma(H) = \sigma(H) - \sigma(0)$ of the five $Bi_{2-x}Cr_xSe_3$ films measured at varied temperatures. For pure $Bi_2Se_3$ (Fig. 2f), the weak field MC curve measured at $T = 1.5K$ shows a sharp negative cusp characteristic of WAL, which has been observed in various TI systems[10-12,23,24]. WAL in TIs reflects the nontrivial topology of the surface states, in which the Dirac fermions travelling along two time-reversed self-crossing (TR-SC) loops accumulate a $\pi$ Berry phase due to spin-momentum locking[25]. The destructive quantum interference between them reduces the return probability, hence the localization tendency of the Dirac fermions, leading to a quantum enhancement to the classical Drude conductance. When a magnetic field is applied perpendicular to the film, the WAL is suppressed. A decrease in conductance, hence a negative MC, is observed.

With the introduction of magnetic impurities, the localization properties of the Dirac fermions are significantly altered. For the lightly doped $x = 0.04$ sample (Fig. 2g), the MC is still negative, but the weak field cusp is not as sharp as that measured at the same $T$ in pure $Bi_2Se_3$, indicating weakened WAL effect. In the intermediately doped $x = 0.07$ sample (Fig. 2h), a qualitatively new feature emerges. The weak field cusp totally vanishes in the curve taken at $T = 1.5K$. MC remains small and negative up to about 2000 Gauss, becomes positive above that, reaches a maximum at around 2 Tesla, and then starts to decrease at even higher field. The non-monotonic behavior disappears as $T$ is warmed to 3K and above, where the MC becomes negative again like those in the less doped samples. When the doping level reaches $x = 0.13$ (Fig. 2i), the $T = 1.5K$ curve shows a sharp positive cusp at weak field, in stark contrast to the WAL behavior in pure $Bi_2Se_3$. This is characteristic of the commonly observed WL, which is the quantum correction to conductance in an opposite manner to WAL due to the constructive quantum interference between the two TR-SC loops. However, as $T$ is warmed slightly to 2K, the weak field MC recovers the negative WAL behavior, although it bends over to WL again at higher field. Both the WAL-like negative MC and the crossover field scale increase with $T$, indicating strengthened WAL by thermal energy. The WL behavior becomes more dominant in the heavily doped $x = 0.23$ sample. Here the WL cusp is extremely sharp at $T = 1.5$ K and persists to $T = 3K$. At higher $T$ the crossover to WAL occurs but the WL behavior is recovered at significantly lower fields than in the $x = 0.13$ sample, indicating much stronger tendency towards WL.

The black symbols in Fig. 3c summarize the $\delta\sigma(H)$ for the five $Bi_{2-x}Cr_xSe_3$ films measured at $T = 1.5K$, which clearly reveal a systematic evolution from typical WAL in $x = 0$ to typical WL in $x = 0.23$. In the intermediate regime, MC shows a competition of these two effects and the total quantum correction to conductivity is suppressed. The black symbols in

Fig. 3d are several representative MC curves in the heavily doped $x = 0.23$ film to highlight the peculiar WL to WAL crossover driven by temperature and magnetic field. Although the patterns shown here are quite rich and complex, the general trend is very clear. Charge transport of the topological surface states exhibits a characteristic crossover from WAL to WL induced by magnetic impurities. The WL behavior becomes more pronounced in the regime of high impurity levels, low temperatures, and strong magnetic fields. Taken collectively, all these factors are favorable for enhancing the perpendicular-to-plane magnetization of the doped impurities. Therefore WL of the topological surface states is intimately related to the degree of TRS breaking by perpendicular magnetization. Whenever the magnetization is weak WAL will be restored by the robust topological protection. The WAL to WL crossover thus results from the competition between topological protection and TRS breaking in the TIs.

The underlying physical picture can be understood intuitively by using the Berry phase language. As discussed above and shown schematically in Fig. 4a, WAL in the pristine topological surface state originates from the helical spin texture of the Dirac fermions, in which the two TR-SC loops accumulate a $\pi$ Berry phase. The perpendicular magnetization of the magnetic impurities tilts the spins of the Dirac fermions out of the plane of motion (Fig. 4b). The Berry phase enclosed by a closed trajectory in the $k$-space thus deviates from $\pi$, causing a less perfect destructive interference and weaker WAL effect. When the spins of Dirac fermions are fully polarized, the Berry phase becomes $2\pi$ and WL is induced by perfect constructive quantum interference. In the intermediate regime with the Berry phase between $\pi$ and $2\pi$, a crossover between WAL and WL naturally takes place.

The role of Berry phase on the localization properties of the topological surface states has been discussed theoretically[26,27]. In particular, it is shown in ref. 27 that in the presence of magnetic impurities the Berry phase can be expressed as

$$\delta\phi = -i\int_0^{2\pi} d\varphi \left\langle \psi_k(r) | \frac{\partial}{\partial\varphi} \psi_k(r) \right\rangle = \pi(1 + \frac{\Delta}{2E_F}). \qquad (1)$$

Here $\Delta$ is the gap opened by magnetic impurities and $E_F$ is the Fermi level measured from the middle point of the gap, as shown schematically in Fig. 4c. In pure TI without magnetic impurity, $\Delta = 0$ and $\delta\phi = \pi$. In the strongly magnetic regime the gap can reach the maximum value $\Delta = 2E_F$ so that we have $\delta\phi = 2\pi$. In the intermediate gap value $\delta\phi$ lies in between $\pi$ and $2\pi$, corresponding to the crossover between WAL and WL.

The evolution of electronic structure with Cr doping has been revealed by ARPES measurements on the $Bi_{2-x}Cr_xSe_3$ films. The ARPES spectrum of pure $Bi_2Se_3$ (Fig. 4d) has well defined Dirac-like surface states and quantum well states. The small gap at the Dirac point is due to the hybridization between the top and bottom surfaces[28]. With increasing Cr concentration, the surface state features become weaker and the gap at the Dirac point becomes wider (Fig. 4d and 4f). In the heavily doped $x = 0.23$ film, the surface states become barely visible, indicating the severe suppression of topological surface states by strong magnetic impurities. Another effect of Cr doping is the systematic shift of $E_F$ towards the bulk valence band, presumably due to the introduction of hole-type carriers. The larger $\Delta$ and smaller $E_F$ in the heavily doped regime causes greater deviation of the Berry phase from $\pi$, leading to stronger localization tendency.

The intricate temperature evolution of the localization behavior in the strongly magnetic sample (Fig. 3d) reveals another aspect of the same physics. At low $T$, the doped magnetic moments may form ferromagnetic ordering, as indicated by the pronounced AHE (Fig. 2e). At high *T*, thermal fluctuation randomizes the orientation of the moments, whereas applied magnetic field tends to align them. The total magnetization, hence the TRS-breaking gap and the Berry phase, is determined by the competition between the thermal energy and Zeeman energy. In the low *T* and strong *H* regime, the magnetic moments are aligned perpendicular to the plane and the system exhibits WL. On the contrary, WAL reenters in the high *T* and weak *H* regime when the magnetic moments become more disordered.

In addition to the phenomenological arguments based on the Berry phase, the quantitative quantum correction to conductance by magnetic impurities was also investigated theoretically[27]. Using the diagrammatic technique by including special diagrams for the TI surface state, it is found that magnetically doped TIs exhibit a unique crossover from WAL to WL via an intermediate regime of classical behavior[27]. The total MC is the result of the competition between WAL and WL, and can be modeled as

$$\delta\sigma(B) = \sum_{i=0,1} \frac{\alpha_i e^2}{\pi h} \left[ \Psi(\frac{l_B^2}{l_\phi^2} + \frac{l_B^2}{l_i^2} + \frac{1}{2}) - \ln(\frac{l_B^2}{l_\phi^2} + \frac{l_B^2}{l_i^2}) \right]. \qquad (2)$$

Here $\Psi$ is the digamma function; the coefficients $\alpha_0$ and $\alpha_1$ stand for the strength of WL and WAL. The parameters $l_B$ is the magnetic length; $l_0$ and $l_1$ are the characteristic length scale of WL and WAL, respectively.

The red curves in Fig. 3c plots the fit of the five MC curves using Eq. 2. We found that for the pure $Bi_2Se_3$ film, WAL dominates so that $\alpha_0$ value is nearly 0 and $\alpha_1$ equals -0.40. In the heavily doped *x* = 0.23 film, WL dominates so that the curve can be fit well by using $\alpha_0$ = 0.14 and a nearly zero $\alpha_1$. For the film in the intermediate regime, especially the *x* = 0.07 sample, the curve can be fit by using very small $\alpha_0$ and $\alpha_1$ values, indicating small quantum corrections to conductance. The open circles in Fig. 3a summarize the evolution of $|α_0|−|α_1|$, the difference between the strength of WL and WAL, with the magnetic impurity level. It increases with $x$ up to $x = 0.13$, following a similar trend as $R_{yx}^0$, but drops in the heavily doped $x = 0.23$ sample. The rough agreement between the two different types of measurements confirms the close relationship between magnetization and localization.

The temperature evolution of MC in magnetically doped TIs was not discussed theoretically, but can be fit by Eq. 2 as well (Fig. 3d) because the same physics dictates the crossover behavior. Again, we find that at low $T$ where WL dominates, the curves can be fit well by using the WL term alone. As $T$ rises, a WAL term with finite $α_1$ has to be included to describe the highly non-monotonic behavior. The open circles in Fig. 3b summarize the temperature dependence of $|α_0|−|α_1|$, which roughly follow the behavior of $R_{yx}^0$, again demonstrating the close ties between magnetization and localization.

We note that the analysis discussed above is oversimplified by ignoring some complications in real materials, such as the hybridization between the two surfaces, possible Kondo-like electron-impurity interaction, and electron-electron interaction. Nevertheless, the systematic trend explained by the Berry phase picture and the excellent fit using the MC formula specially derived for TIs demonstrate that the unique WL to WAL crossover reflects the intrinsic physics of the topological surface states under TRS-breaking perturbations. The tuning of topological transport properties by magnetic doping, temperature, and magnetic field may pave the road for investigating the predicted exotic effects and possible spintronic applications created by the intricate interplay between topological protection and broken TRS in TIs.

## Method summary

The MBE growth of $Bi_2Se_3$ ultrathin films on sapphire(0001) substrate has been reported elsewhere[29]. In this work the Cr dopants are evaporated simultaneously with the Bi and Se atoms. The Cr concentration $x$ is controlled by the evaporation temperature (from 960ºC for $x$ = 0.04 to 1020ºC for $x$ = 0.23) of the Knudsen cell containing the Cr source. The $x$ value is determined by an atomic emission spectrometer on 100QL thick $Bi_{2-x}Cr_xSe_3$ films grown under the same condition as the 3QL films used here. Transport properties are measured by standard four-probe ac lock-in method. ARPES measurements are performed *in situ* at $T \sim$ 100K by using a Scienta SES-2002 analyzer. All the ARPES spectra are taken along the Γ-K direction.

## Acknowledgements

We thank X. L. Qi, S. Q. Shen, C. S. Tian, Z. Wang, and Z. Y. Weng for helpful discussions. This work was supported by the National Natural Science Foundation of China, the Ministry of Science and Technology of China, and the Chinese Academy of Sciences.

## Competing financial interests

The authors declare that they have no competing financial interests.

# References

1. Qi, X. L. & Zhang, S. C. The quantum spin Hall effect and topological insulators. *Phys. Today* **63**, 33-38 (2010).

2. Moore, J. E. The birth of topological insulators. *Nature* **464**, 194-198 (2010).

3. Hasan, M. Z. & Kane, C. L. Topological Insulators. *Rev. Mod. Phys.* **82**, 3045-3067 (2010).

4. Hsieh, D. *et al.* A tunable topological insulator in the spin helical Dirac transport regime. *Nature* **460**, 1101-1105 (2009).

5. Hsieh, D. *et al.* Observation of Unconventional Quantum Spin Textures in Topological Insulators. *Science* **323**, 919-922 (2009).

6. Chen, Y. L. *et al.* Experimental Realization of a Three-Dimensional Topological Insulator, Bi2Te3. *Science* **325**, 178-181 (2009).

7. Roushan, P. *et al.* Topological surface states protected from backscattering by chiral spin texture. *Nature* **460**, 1106-1109 (2009).

8. Zhang, T. *et al.* Experimental Demonstration of Topological Surface States Protected by Time-Reversal Symmetry. *Phys. Rev. Lett.* **103**, 266803 (2009).

9. Alpichshev, Z. *et al.* STM Imaging of Electronic Waves on the Surface of Bi2Te3: Topologically Protected Surface States and Hexagonal Warping Effects. *Phys. Rev. Lett.* **104**, 016401 (2010).

10. Checkelsky, J. G. *et al.* Quantum Interference in Macroscopic Crystals of Nonmetallic Bi2Se3. *Phys. Rev. Lett.* **103**, 246601 (2009).

11. Peng, H. *et al.* Aharonov-Bohm interference in topological insulator nanoribbons. *Nature Mat.* **9**, 225-229 (2010).

12. Chen, J. *et al.* Gate-Voltage Control of Chemical Potential and Weak Antilocalization in Bi2Se3. *Phys. Rev. Lett.* **105**, 176602 (2010).

13. Qi, X. L., Hughes, T. L. & Zhang, S. C. Topological field theory of time-reversal invariant insulators. *Phys. Rev. B* **78**, 195424 (2008).

14. Liu, Q., Liu, C. X., Xu, C. K., Qi, X. L. & Zhang, S. C. Magnetic Impurities on the Surface of a Topological Insulator. *Phys. Rev. Lett.* **2009**, 156603 (2008).

15. Qi, X. L., Li, R., Zang, J. & Zhang, S. C. Inducing a Magnetic Monopole with Topological Surface States. *Science* **323**, 1184-1187 (2009).

16. Fu, L. & Kane, C. L. Probing Neutral Majorana Fermion Edge Modes with Charge Transport. *Phys. Rev. Lett.* **102**, 216403 (2009).

17. Yu, R. *et al.* Quantized Anomalous Hall Effect in Magnetic Topological Insulators. *Science* **329**, 61-64 (2010).

18. Tse, W. K. & MacDonald, A. H. Giant Magneto-Optical Kerr Effect and Universal Faraday Effect in Thin-Film Topological Insulators. *Phys. Rev. Lett.* **105**, 057401 (2010).

19. Xia, Y. *et al.* Observation of a large-gap topological-insulator class with a single Dirac cone on the surface. *Nature Phys.* **5**, 398-402 (2009).

20. Zhang, H. J. *et al.* Topological insulators in Bi2Se3, Bi2Te3 and Sb2Te3 with a single Dirac cone on the surface. *Nature Phys.* **5**, 438-442 (2009).

21. Chen, Y. L. *et al.* Massive Dirac Fermion on the Surface of a Magnetically Doped Topological Insulator. *Science* **329**, 659-662 (2010).

22. Wray, L. A. *et al.* A topological insulator surface under strong Coulomb, magnetic and disorder perturbations. *Nature Phys.* **7**, 32-37 (2010).

23  Liu, M. H. *et al.* Electron interaction-driven insulating ground state in Bi2Se3 topological insulators in the two dimensional limit. *to appear in Phys. Rev. B* (2011).

24  He, H. T. *et al.* Impurity effect on weak anti-localization in topological insulator Bi2Te3. arxiv:1008.0141v1001 (2010).

25  Nomura, K., Koshino, M. & Ryu, S. Topological Delocalization of Two-Dimensional Massless Dirac Fermions. *Phys. Rev. Lett.* **99**, 146806 (2007).

26  Ghaemi, P., Mong, R. S. K. & Moore, J. E. In-Plane Transport and Enhanced Thermoelectric Performance in Thin Films of the Topological Insulators Bi2Te3 and Bi2Se3. *Phys. Rev. Lett.* **105**, 166603 (2010).

27  Lu, H. Z., Shi, J. R. & Shen, S. Q. Competing weak localization and anti-localization in topological surface states. arXiv:1101.5437 (2011).

28  Zhang, Y. *et al.* Crossover of the three-dimensional topological insulator Bi2Se3 to the two-dimensional limit. *Nature Phys.* **6**, 584-588 (2010).

29  Chang, C. Z. *et al.* Growth of quantum well films of topological insulator Bi2Se3 on insulating substrate arXiv:1012.5716 (2010).

Figure Captions

**Figure 1 | Schematic setup and the resistance vs temperature curves. a,** Schematic of the 3QL thick $Bi_{2-x}Cr_xSe_3$ films grown on sapphire for the transport studies. The thickness is not to scale. In all the magneto transport measurements the field orientation is perpendicular to the film. **b,** The $R - T$ curves of the five $Bi_{2-x}Cr_xSe_3$ films all show metallic behavior at high $T$ but becomes insulating at low $T$. **c,** Both the minimum resistance value ($R_{min}$) and the onset temperature ($T_{onset}$) for the insulating behavior increases with $x$, indicating the weakening of electron conduction capability of the topological surface states by magnetic impurities.

**Figure 2 | Field dependence of the Hall effect and magnetoconductance (MC) in $Bi_{2-x}Cr_xSe_3$ films. a,** The pure ($x = 0$) $Bi_2Se_3$ film shows a linear Hall effect. **b, c,** and **d,** In Cr doped films ($x = 0.04, 0.07$, and $0.13$) the nonlinear Hall effect starts to appear and the nonlinearity becomes more pronounced with increasing $x$. **e,** The most heavily doped ($x = 0.23$) film has strongly nonlinear Hall curves up to 20K. The broken line shows how the intercept $R_{yx}^0$ is obtained for the $T = 1.5K$ curve. **f,** MC of the pure ($x = 0$) $Bi_2Se_3$ film shows a negative cusp characteristic of WAL. **g,** Lightly doped ($x = 0.04$) film shows WAL with reduced magnitude and broader field scale than the pure sample. **h,** MC of the $x = 0.07$ film shows a non-monotonic behavior at $T = 1.5K$. At slightly higher $T$ the WAL behavior reenters. **i,** The $x = 0.13$ film shows quite sharp WL-like cusp at $T = 1.5K$. At higher $T$, WAL dominates at low fields while WL is recovered at high fields. **j,** The most heavily doped ($x = 0.23$) film shows qualitatively similar behavior as the $x = 0.13$ sample but here the WL behavior becomes more prevalent.

**Figure 3 | Correlation between magnetization and localization. a,** Red solid squares show the low temperature ($T = 1.5K$) $R_{yx}^0$, which is roughly proportional to the perpendicular magnetization, as a function of Cr doping level. Black open circles show $|\alpha_0|-|\alpha_1|$, the difference between the strength of WL and WAL obtained by the data fitting in **c**. Comparison with $R_{yx}^0$ demonstrates that larger magnetization tends to enhance WL. **b,** Red

solid squares show $T$ dependence of the $R_{yx}^{0}$ for the most heavily doped $x = 0.23$ film. Black open circles show $|\alpha_0|-|\alpha_1|$, again revealing the correlation between magnetization and localization. **c,** Black symbols show the low temperature ($T = 1.5K$) MC for all five $Bi_{2-x}Cr_xSe_3$ films. The $x = 0$ and $x = 0.23$ curves are scaled by a factor of 0.5 and 3 respectively for clarity. A crossover from WAL to WL is observed as $x$ increases. Red lines are the fits of the MC curves using Eq. 2. **d,** Black symbols show the MC curves at several representative $T$ for the $x = 0.23$ sample to show the $T$ and $H$ evolution of the localization behavior. Red lines are the fits of the MC curves using Eq. 2.

**Figure 4 | Berry phase and the TRS-breaking gap. a,** Schematic illustration of the quantum interference between two TR-SC loops of the spin-momentum-locked topological surface state. **b,** The perpendicular magnetization tilts the spins of the Dirac fermions out of the plane of motion, causing a deviation of the Berry phase from π. **c,** The magnetization breaks the TRS and opens a gap at the Dirac point of the surface state. **d,** ARPES spectrum for the pure $Bi_2Se_3$ film ($x = 0$). A Dirac-like dispersion characteristic of the topological surface states is well resolved. A hybridization-induced Δ ~ 100meV can be seen near the Driac point. **e,** ARPES data for the intermediately doped $x = 0.07$ film, where the surface state feature becomes weaker and the gap amplitude becomes larger. **f,** The gap amplitude continues to increase in the strongly magnetic $x = 0.23$ film, and the surface state feature becomes extremely weak due to the severe suppression of topological surfaces by strong magnetic impurities.

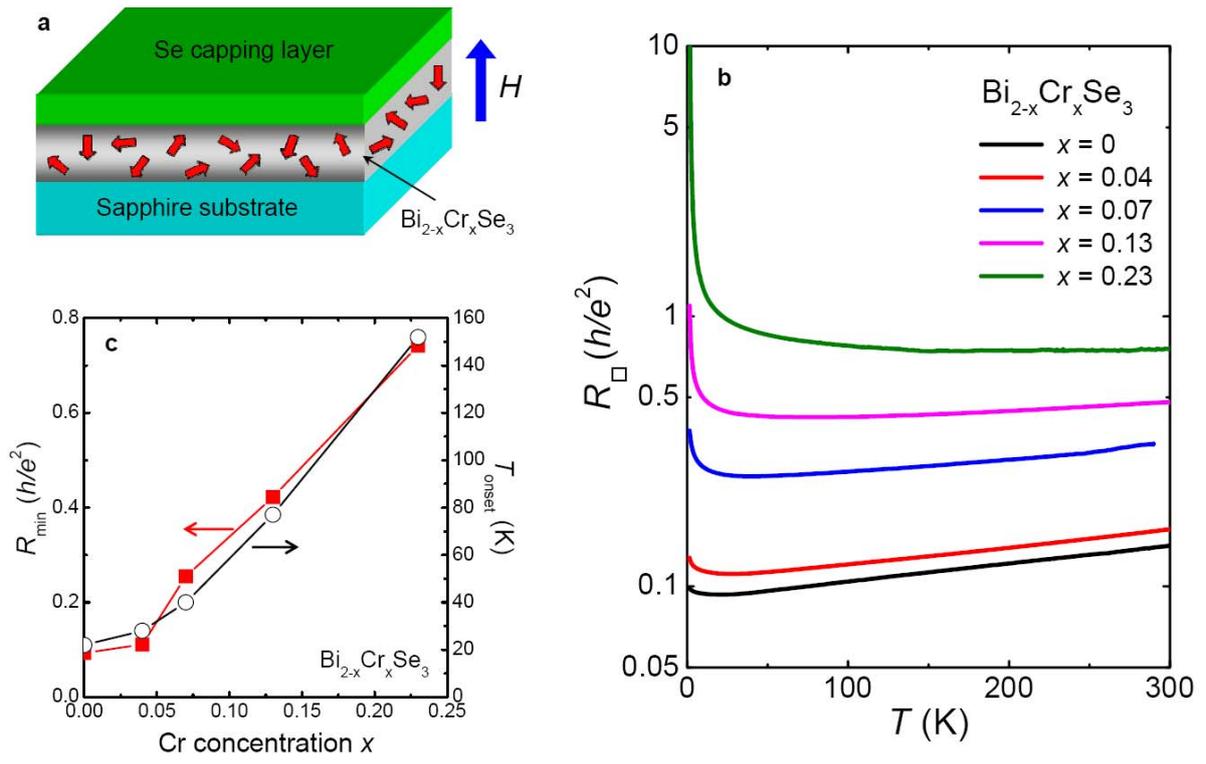

Figure 1

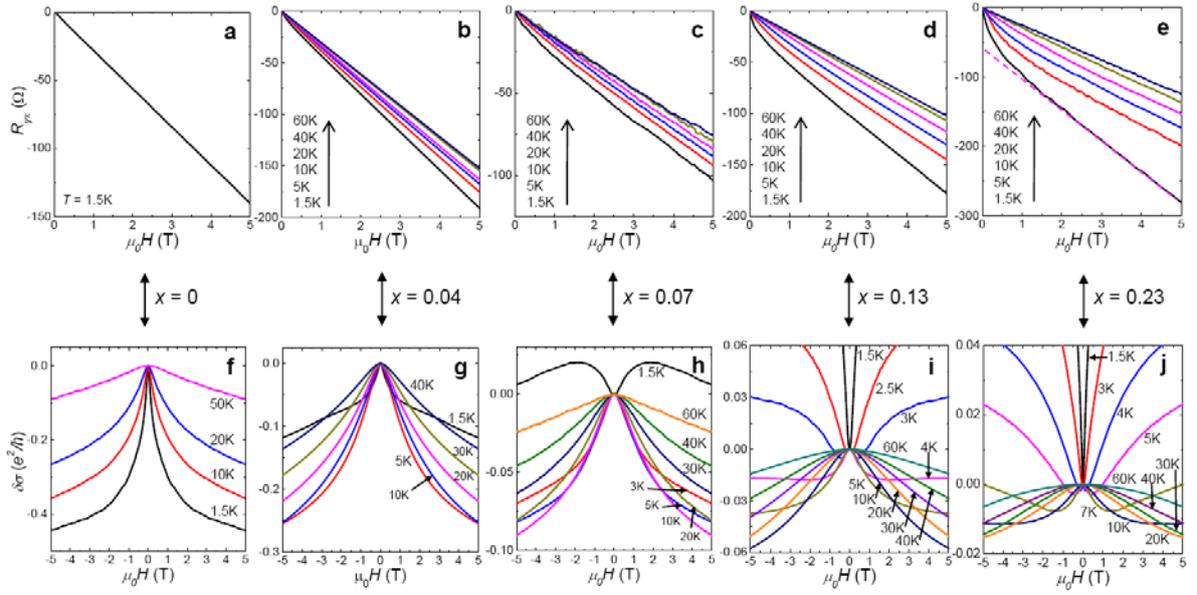

Figure 2

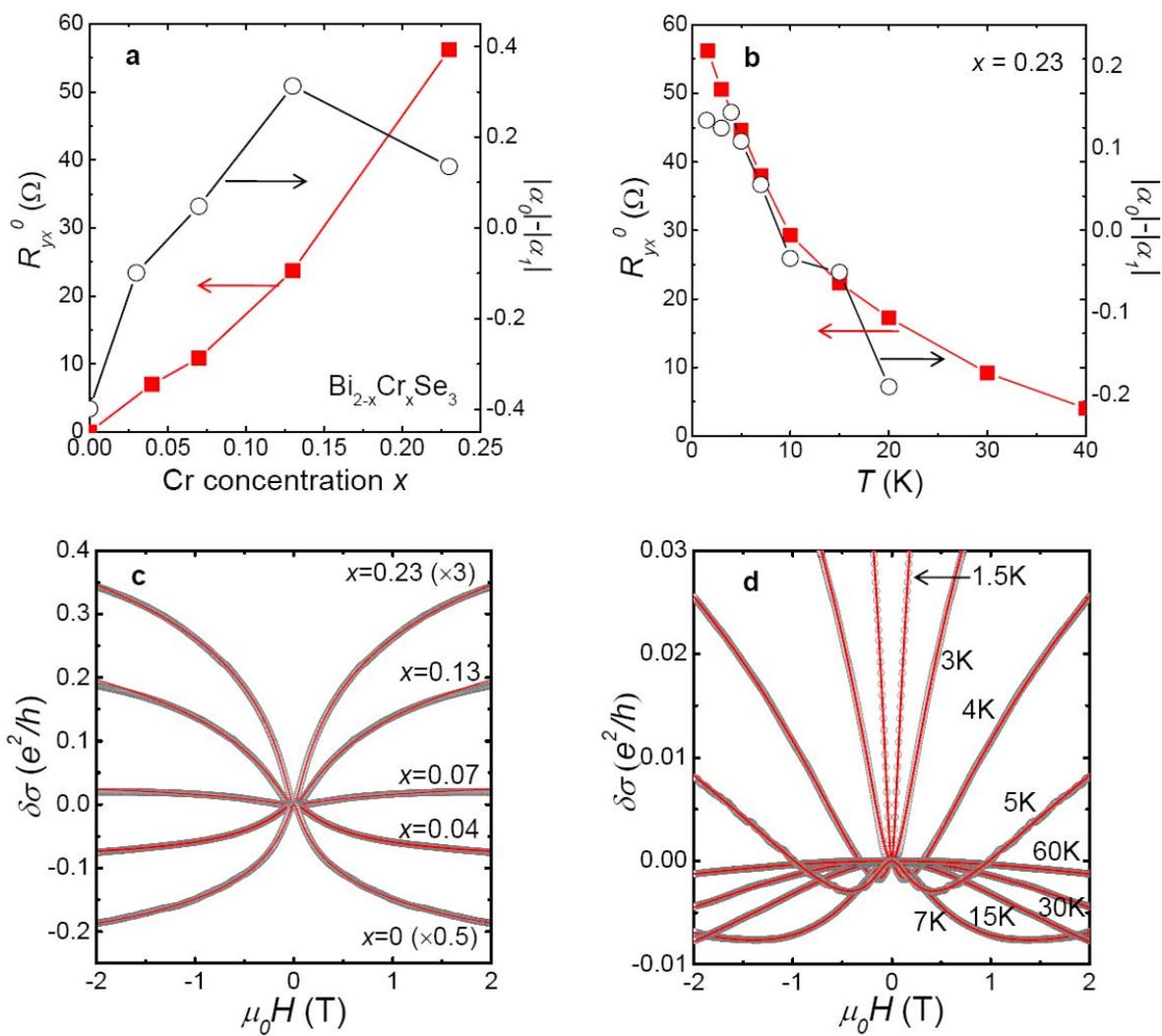

Figure 3

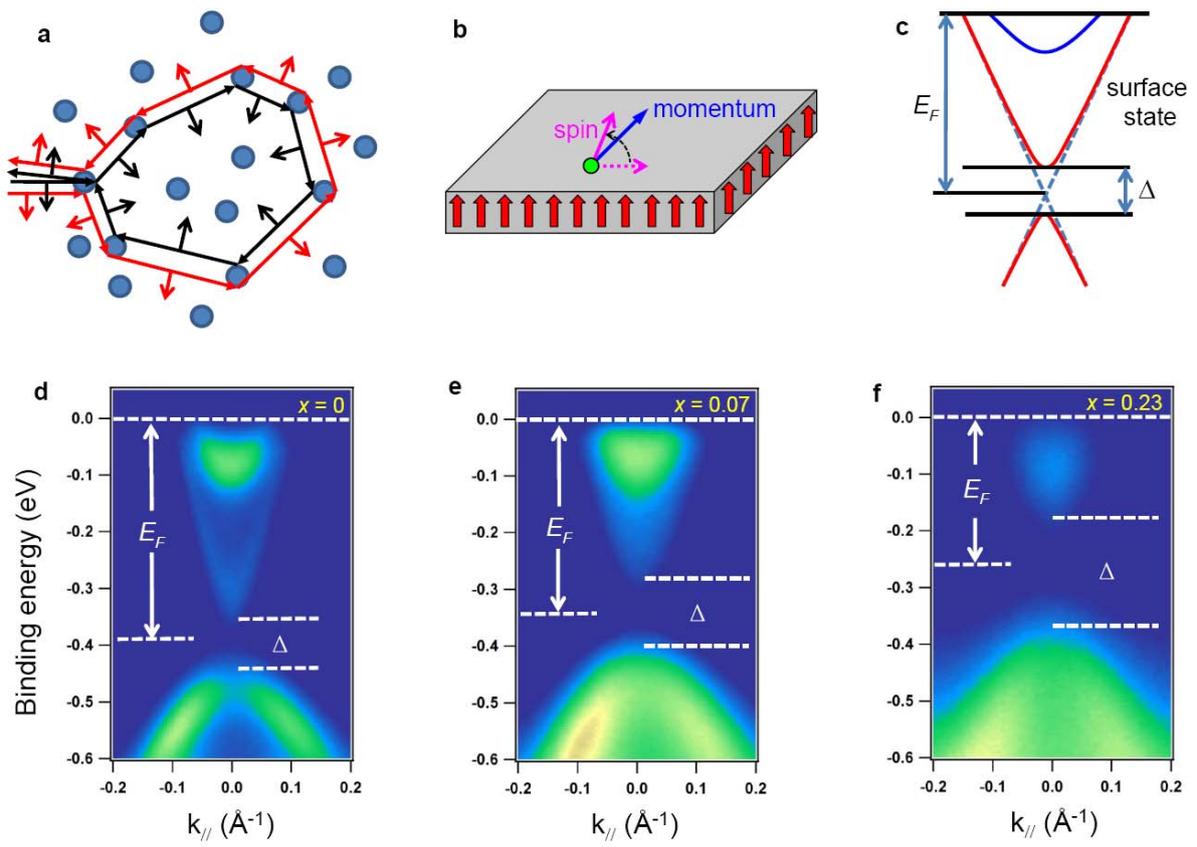

Figure 4